\documentclass[aps,prb,twocolumn,groupedaddress,showpacs]{revtex4}
\usepackage{graphicx}
\usepackage{dcolumn}
\usepackage{bm}
\usepackage{amsmath}

\begin{document}


\title{Spin-orbit coupling in ferromagnetic Nickel} 
 
\author{J.~B\"unemann$^1$, F.~Gebhard$^1$, T.~Ohm$^2$, S.~Weiser$^2$, and 
W.~Weber$^2$}
\affiliation{\centerline{$^{1}$
Fachbereich Physik and Material Sciences Center, Philipps--Universit\"at Marburg, 
D--35032 Marburg, Germany}\\
$^{2}$Fachbereich Physik,  Universit\"at Dortmund, D--44221 Dortmund,
 Germany}
                 
\begin{abstract}%
We use the Gutzwiller variational theory to investigate the 
electronic and the magnetic properties of fcc-Nickel. Our particular focus 
 is on the effects of the spin-orbit coupling. Unlike 
 standard relativistic band-structure theories, we  
 reproduce the experimental magnetic moment direction
and we explain the change 
of the Fermi-surface topology that occurs when the magnetic moment 
 direction is rotated by an external magnetic field.
 The Fermi surface in our calculation deviates from early 
 de-Haas--van-Alphen (dHvA) results. We attribute these discrepancies to an 
 incorrect interpretation of the raw dHvA data.   
\end{abstract}
          
\maketitle 
The limitations of density functional theory (DFT) when treating 
 the  electronic and magnetic properties of transition metals
 become evident most clearly in the case of Nickel. The DFT cannot 
 reproduce gross features such as the width of the $3d$-bands 
($4.5 \,{\rm eV}$ versus $3.3 \,{\rm eV}$ 
experimentally\cite{ehk78,ep80,moruzzi}),
 nor important details  such as the exchange splitting. 
The exchange splitting in the DFT
 is almost $0.7\,{\rm eV}$ and rather isotropic over the Fermi-surface,
 whereas, experimentally, it is found to be much smaller and strongly 
 orbital dependent: $\Delta_{e_{g}}\approx0.17\,{\rm eV}$ and 
$\Delta_{t_{2g}}\approx0.33\,{\rm eV}$. As a result, even the Fermi 
 surface topologies do not match, because of the position of the 
  $X_{2,\downarrow}$ energy: above $E_{\rm F}$ in DFT, yet below $E_{\rm F}$
 experimentally; thus only one hole ellipsoid exists around the 
 $X$ point, versus two in DFT.\cite{epl,tsui}
 
Even more limitations of DFT become evident when the effects of the 
spin-orbit coupling are considered. The magnetic anisotropy energy has 
the wrong sign for Nickel (and for Cobalt), while it has the correct 
sign for Iron, yet is too small by a factor of three. \cite{daalderop}
In Nickel, the easy axis is along $[111]$ 
 and approximately $3\,\mu {\rm eV}$ per atom are needed to rotate the 
 magnetic moment axis into the $[001]$ direction.\cite{gersdorf,gersdorf2}  
Moreover, a detailed 
 low-temperature study  of the magnetic anisotropy constants 
 $K_1,K_2,K_3$  by Gersdorf \cite{gersdorf2} has revealed a change in 
the Fermi-surface topology
 when the magnetic-moment axis is rotated into the $[001]$ direction: 
 A small second hole ellipsoid appears around the $X(001)$ point, but not 
around the $X(100)$ and  $X(010)$ points, now inequivalent to $X(001)$, 
because  of the underlying tetragonal symmetry.

It is generally accepted that the discrepancies between the 
DFT and the experimental results are mainly 
caused by an insufficient treatment of the electronic correlation 
in an effective one-particle theory. 
 In the past, all attempts to combine the DFT with more sophisticated 
 correlated electron theories have only led to partial improvements of the 
results  for Nickel; see, e.g., the GW approximation in Ref. 
\onlinecite{ary92}.

 In a recent work\cite{epl}  we were able to show that a generalized 
Gutzwiller theory provides a consistent picture of the quasi-particle 
band-structure of Nickel. Neglecting spin-orbit coupling,
 all basic problems of the DFT calculations on Nickel  have been resolved.
Our theory employed approximately $2^{10}$ variational parameters 
 representing the occupancies of all atomic multi-electron states within an 
open $3d$ shell (see below).  

In this letter we present results for the case 
when spin-orbit coupling is included. In order
 to cope with this complication, the Gutzwiller theory had to be 
 extended \cite{narlikar} to allow for rotations in the eigenvector space 
 of the atomic atomic multi-electron states, resulting in many 
 more variational parameters. Employing this generalization 
 we obtain the correct magnetic anisotropy energy, and, more importantly, 
reproduce the change in the Fermi-surface topology found by Gersdorf.

To investigate transition metals we start from multi-band Hubbard models 
of the general form  
\begin{equation}\label{1.1}
\hat{H}=\sum_{i \ne j;\sigma,\sigma'}t_{i,j}^{\sigma,\sigma'}\hat{c}_{i,\sigma}^{\dagger}\hat{c}_{j,\sigma'}^{\phantom{+}}
+\sum_i \hat{H}_{{\rm loc},i}=\hat{H}_0+\hat{H}_{\rm loc}\;.
\end{equation}
 Here, the first term describes the hopping of electrons between spin-orbital 
states $\sigma,\sigma'$ on lattice sites $i,j$, respectively. The Hamiltonian
 \begin{equation}\label{1.1b}
\hat{H}_{{\rm loc},i}=\hat{H}_{{\rm C},i}+\hat{H}_{{\rm cf},i}+\hat{H}_{{\rm SO},i}
\end{equation}
contains all local terms, i.e., the two-particle Coulomb 
interaction $\hat{H}_{{\rm C},i}$, the crystal field energies $
\hat{H}_{{\rm cf},i}$ and the spin-orbit coupling 
$\hat{H}_{{\rm SO},i}$. In the case of Nickel, we work with 
 a basis of $3d$, $4s$, and $4p$ orbitals. 

We have determined the bare hopping-parameters in the one-particle Hamiltonian
 $\hat{H}_0$ and the crystal-field energies in $\hat{H}_{\rm cf}$
  by means of a tight-binding fit to the 
paramagnetic DFT band structure.\cite{epl,narlikar}
Due to the large 
 band-width of the $4s$ and $4p$ bands, only the 
Coulomb-interaction within the $3d$-shell is taken into account. 
The spherical approximation is used, i.e., we express 
the Coulomb interaction through the three Racah-parameters
 $A$, $B$, and $C$.\cite{sugano} Note that cubic site symmetry would allow 
ten independent interaction parameters.  
 In order to reproduce the experimental 
$d$-band width in our approach, we need a Racah-parameter 
$A\approx 9 \,{\rm eV}$. The Racah-parameters $B$ and $C$ are assumed to be 
 close to their atomic values \cite{sugano},  $B\approx 85\,{\rm meV}$ and 
$C\approx 400\,{\rm meV}$, resulting in a value $J$
 of $J=7B/2+7C/5\approx 0.85 \; {\rm eV}$.   
The spin-orbit coupling  parameter
 $\zeta$ in the spin-orbit Hamiltonian 
\begin{equation}
\hat{H}_{{\rm SO},i} = \sum_{\sigma\sigma'} \frac{\zeta}{2}
\langle \sigma | \widehat{l}_x \widetilde{\sigma}_x
+ \widehat{l}_y \widetilde{\sigma}_y + \widehat{l}_z \widetilde{\sigma}_z
| \sigma'\rangle
\hat{c}_{i,\sigma}^+\hat{c}_{i,\sigma'} \; 
\label{spinorbitHamilt}
\end{equation}
is chosen as $\zeta= 80 {\rm meV}$. Note that the Hamiltonian 
 (\ref{spinorbitHamilt}) only contains $d$-orbitals. 

In the Gutzwiller theory, the following Ansatz for a variational wave-function 
 \cite{gutzwiller,prb98,narlikar}
 is used to investigate  the multi-band Hubbard model (\ref{1.1})
\begin{equation}\label{1.3}
|\Psi_{\rm G}\rangle=\hat{P}_{\rm G}|\Psi_0\rangle=\prod_{i}\hat{P}_{i}|\Psi_0\rangle\;.
\end{equation}
Here, $|\Psi_0\rangle$ is a normalized single-particle product state and the 
local Gutzwiller correlator is defined as 
\begin{equation}\label{1.4}
\hat{P}_{i}=\sum_{\Gamma,\Gamma^{\prime}}\lambda^{(i)}_{\Gamma,\Gamma^{\prime}}
|\Gamma \rangle_{i} {}_{i}\langle \Gamma^{\prime} |\;.
\end{equation}
The states $| \Gamma \rangle_i$ form some arbitrary basis of the atomic 
Hilbert-space and the (complex) numbers $\lambda^{(i)}_{\Gamma,\Gamma^{\prime}}$ 
are variational parameters. For Nickel, we work with a correlation 
operator~(\ref{1.4}) in which the states $| \Gamma \rangle_i$ are 
the eigenstates of the atomic Hamiltonian 
$ \hat{H}_{{\rm C},i}$. The non-diagonal elements of the 
variational parameter matrix 
$\lambda_{\Gamma,\Gamma'}$ are assumed to 
be finite only for states $|\Gamma \rangle$, $|\Gamma' \rangle$ which belong 
to the same atomic multiplet. This is consistent with the spherical 
approximation for the Coulomb-interaction. In the case of Nickel, it is 
 sufficient to work with non-diagonal parameters $\lambda_{\Gamma,\Gamma'}$ 
 in the $d^7$, $d^8$ and $d^9$ shells. 
 
The expectation value of the Hamiltonian (\ref{1.1})  can be calculated 
 analytically for the Gutzwiller wave function (\ref{1.3}) in the limit
 of infinite spatial dimensions.\cite{prb98}
We use the exact results in this limit
 as an approximation for our three-dimensional model system. The Gutzwiller 
energy functional can also be obtained in slave-boson mean-field theory.\cite{lechermann,slave}
 In infinite dimensions one finds 
 \begin{equation}\label{1.5}
\langle \hat{H}_{{\rm loc},i} \rangle_{\Psi_{\rm G}}=\sum_{\Gamma_{1}\ldots\Gamma_{4}}\lambda_{\Gamma_{2},\Gamma_{1}}^{*}\lambda_{\Gamma_{3},\Gamma_{4}}^{}
E^{\rm loc}_{\Gamma_{2},\Gamma_{3}}m^0_{\Gamma_1,\Gamma_4}
\end{equation}
for the expectation value of the local Hamiltonian in (\ref{1.1}),
where 
\begin{eqnarray}\label{1.6}
E^{\rm loc}_{\Gamma_{2},\Gamma_{3}}&\equiv&\langle 
\Gamma_{2} | \hat{H}_{{\rm loc},i}  | \Gamma_{3}\rangle\;,\\
\label{1.6b}
m^0_{\Gamma_1,\Gamma_4} &\equiv& \langle \left( |\Gamma_{1} \rangle   \langle  \Gamma_{4} |\right)  \rangle_{\Psi_0}\;.
\end{eqnarray} 
The local expectation value (\ref{1.6b}) is readily calculated by means of
 Wick's theorem. 
For the expectation value of a hopping operator in (\ref{1.1}) one finds
\begin{equation}\label{1.13a}
\langle  \hat{c}_{i,\sigma_1}^{\dagger}\hat{c}_{j,\sigma_2}^{\phantom{+}} \rangle_{\Psi_{\rm G}}=\sum_{\sigma'_1,\sigma'_2}q_{\sigma_1}^{\sigma'_1}\left( q_{\sigma_2}^{\sigma'_2}\right)^{*}\langle  
\hat{c}_{i,\sigma'_1}^{\dagger}\hat{c}_{j,\sigma'_2}^{\phantom{+}} \rangle_{\Psi_{0}}\;.
\end{equation}
The renormalization matrix 
$q_{\sigma}^{\sigma'}$ in  (\ref{1.13a}) 
can be calculated most easily when an orbital basis is used which has a 
 diagonal local density-matrix with respect to $|\Psi_{0}\rangle$,
\begin{equation}\label{1.14}
C^0_{\sigma,\sigma'}\equiv\langle \hat{c}_{i,\sigma}^{\dagger}  \hat{c}_{i,\sigma'}^{} \rangle_{\Psi_{0}}
=\delta_{\sigma,\sigma'}n^0_{\sigma}.
\end{equation} 
If $C^0_{\sigma,\sigma'}$ is non-diagonal for a one-particle product state 
$|\Psi_{0}\rangle$ one can always transform the orbital basis in order to
 ensure that Eq.~(\ref{1.14}) holds.
In the case of a diagonal local 
 density-matrix~(\ref{1.14}), the renormalization matrix 
in (\ref{1.13a}) reads
 \begin{eqnarray}\label{1.15}
q_{\sigma}^{\sigma'}=\frac{1}{n^{0}_{\sigma'}}
\sum_{\Gamma_1\ldots\Gamma_4}\lambda^{*}_{\Gamma_2,\Gamma_1}
\lambda^{}_{\Gamma_3,\Gamma_4}
\langle \Gamma_2|
\hat{c}_{\sigma}^{\dagger}
|\Gamma_3\rangle&&\\ \nonumber
\times \left \langle
\left(|\Gamma_1  \rangle
\langle \Gamma_4 |  \hat{c}_{\sigma'}
  \right)
\right \rangle_{\Psi_0}&& \;,
\end{eqnarray} 
where, again, the expectation value with respect to $|\Psi_0 \rangle$ is
 calculated with Wick's theorem. 

The variational ground-state energy must be minimized with respect to 
 the variational parameters $\lambda_{\Gamma,\Gamma'}$ and the 
  one-particle product wave-functions $|\Psi_0 \rangle$. It has been found 
\cite{narlikar,thul} that the optimum state $|\Psi_0 \rangle$ is the 
 ground state of the effective one-particle Hamiltonian 
\begin{equation}
\hat{H}_0^{\rm eff}=\sum_{i \ne j;\sigma,\sigma'}\tilde{t}_{i,j}^{\sigma,\sigma'}
\hat{c}_{i,\sigma}^{\dagger}\hat{c}_{j,\sigma'}^{\phantom{+}}+
\sum_{i;\sigma,\sigma'}\eta_{\sigma,\sigma'}
\hat{c}_{i,\sigma}^{\dagger}\hat{c}_{i,\sigma'}^{\phantom{+}}\,.
\end{equation}
Here, we introduced the renormalized hopping matrix elements
\begin{equation}
\tilde{t}_{i,j}^{\sigma_1,\sigma_2}=\sum_{\sigma'_1,\sigma'_2}
q_{\sigma'_1}^{\sigma_1}\left( q_{\sigma'_2}^{\sigma_2}\right)^{*}
t_{i,j}^{\sigma'_1,\sigma'_2}
\end{equation}
and the Lagrange-parameters $\eta_{\sigma,\sigma'}$ which are used 
 to optimize the energy with respect to the local density matrix 
 (\ref{1.14}). Within a Landau Fermi-liquid approach one can further 
show\cite{narlikar,thul} that  the eigenvalues 
$E_{\gamma}(k)$ of $\hat{H}_0^{\rm eff}$ are 
the quasi-particle excitation energies that can be compared, for example, 
to  ARPES experiments. Most important for the quasi-particle 
band-structure are the Lagrange-parameters $\eta^{d}_{\sigma,\sigma'}$ 
for the $d$-orbitals. The two (diagonal) Lagrange parameters for the 
$s$ and $p$  orbitals are adjusted in order to fix 
the total $d$-electron number.\cite{comment1} 

The inclusion of spin-orbit coupling in the Gutzwiller theory complicates 
the numerical  minimization significantly. Both the $d$-part of the 
local density-matrix (\ref{1.14}) and of the hopping renormalization 
matrix  (\ref{1.15}) are no longer diagonal.  The number $n_{\rm ie}$ of 
independent elements depends on the magnetic moment direction, 
we find $n_{\rm ie}=22$ for   $\vec{\mu}||[111]$ and  $n_{\rm ie}=18$ 
for $\vec{\mu}||[001]$.
 As a consequence of the reduced symmetry, we could work
  with up to $n_{\rm ie}$ independent $d$-shell Lagrange parameters 
$\eta^{d}_{\sigma,\sigma'}$ 
 in order to minimize the total energy. Numerically, however, such a 
minimization would be quite costly since each variation of these parameters 
 involves many momentum-space integrations. We therefore work with a simplified
effective Hamiltonian $\hat{\tilde{H}}^{\rm eff}_{0}$ that contains effective 
parameters only for all physically relevant one-particle terms.

In cubic symmetry, there exist only four independent matrix 
 elements of the local ($d$-electron) density-matrix. The trace
 of the matrix is fixed by the total $d$-electron number. The three 
 remaining matrix elements are governed by parameters 
$\eta^{d}_{\sigma,\sigma'}$ which are given by the orbital-dependent 
exchange fields $\Delta_{t_{2 {\rm g}}}$, 
$\Delta_{e_{{\rm g}}}$ and the effective crystal-field splitting 
$ \epsilon^{\rm eff}_{\rm CF}$.

The non-cubic symmetry resulting from the addition of the spin-orbit 
 coupling adds many more formally independent
$\eta^{d}_{\sigma,\sigma'}$ terms. Both for $\vec{\mu}||[001]$ 
(tetragonal symmetry) and for $\vec{\mu}||[111]$ (trigonal symmetry)
 there are two more  exchange-fields and two more crystal-field splittings. 
All these eight terms are included in our simplified
effective Hamiltonian $\hat{\tilde{H}}^{\rm eff}_{0}$.
In the spirit of the spherical approximation\cite{callaway}, 
 a Hamiltonian $\hat{H}^{\rm eff}_{\rm SO}$ is included in 
$\hat{\tilde{H}}^{\rm eff}_{0}$ that 
 has the same form as $\hat{H}_{\rm SO}$ only with  $\zeta$ replaced
 by $\zeta^{\rm eff}$. As a result, we have to minimize the total energy 
with respect to nine `external' parameters in our simplified Hamiltonian 
$\hat{\tilde{H}}^{\rm eff}_{0}$.

The numerical minimization is much more time-con\-suming for a system
 with spin-orbit coupling than without. First, spin-orbit coupling
 requires the momentum-space integration to be extended from $1/48$th to the 
 full Brillouin zone. Furthermore, the small values of the anisotropy energy 
 necessitate a much finer mesh for the momentum-space integration. Second, 
 the energy needs to be minimized with respect to nine external 
 variational parameters in $\hat{\tilde{H}}^{\rm eff}_{0}$.
 Altogether, the minimization of the total energy 
 is approximately $10^4$ times more time consuming for a system with 
spin-orbit coupling than in the absence of $\hat{H}_{\rm SO}$.

 \begin{figure}
\centering
\includegraphics[scale=0.75]{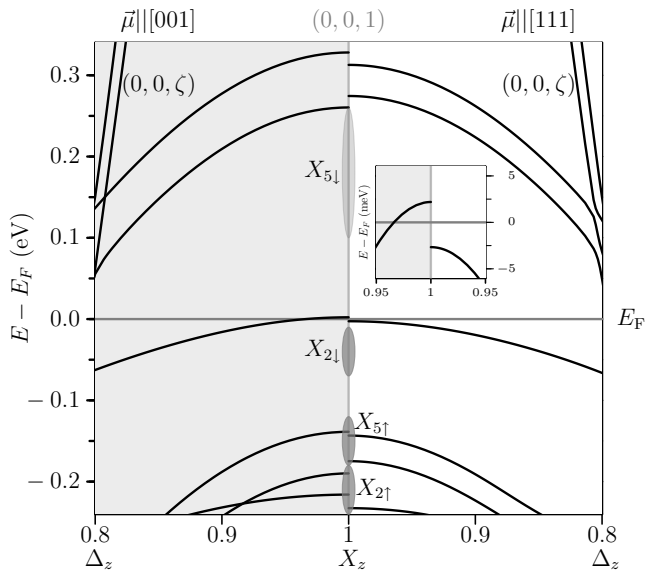}
\caption{Quasi-particle band structure along the $\Delta$-line 
around the $X_z$-point of the Brillouin zone for magnetic-moment
directions $\vec{\mu}\parallel [001]$ and $\vec{\mu}\parallel [001]$.
The inset shows an enlarged view of the band structure
around the Fermi energy which displays the additional hole ellipsoid 
for $\vec{\mu}\parallel [001]$ more clearly. \label{fig1}}
\end{figure} 

 \begin{figure}
\centering
\includegraphics[scale=0.75]{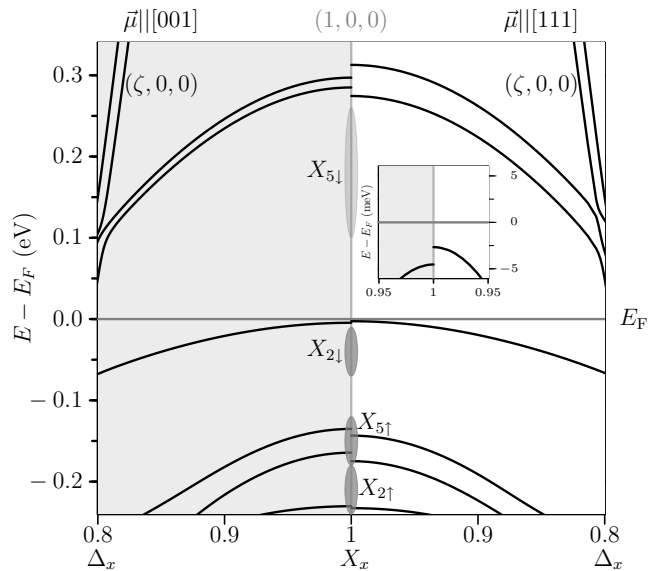}
\caption{Same as in Fig.~\protect\ref{fig1} but around the 
$X_x$-point. Note that the $X_{2\downarrow}$-band is below
the Fermi energy for both magnetic-moment directions. \label{fig2}}
\end{figure}

We carried out the minimization of the variational energy with respect to the 
 `internal' parameters $\lambda_{\Gamma,\Gamma'}$ and the 
external parameters for both magnetic moment directions  
$\vec{\mu}||[111]$ and 
$\vec{\mu}||[001]$. The optimum value of the effective spin-orbit 
 coupling is $\zeta_{\rm eff}\approx 68 \,{\rm meV}$ in both cases, 
about $15\%$ smaller than 
 the bare value $\zeta=80 \,{\rm meV}$. There seems to be no simple rule that 
 determines the relative size of $\zeta$ and  $\zeta_{\rm eff}$.  
 For example, for Iron  we found an effective spin-orbit 
 considerably larger than the corresponding bare value. In our calculations 
 for Nickel, the anisotropy energy is  
$E_{\rm aniso}\approx 3.5 \, \mu {\rm eV}$ per atom, 
 quite close to the experimental value 
$E_{\rm exp}\approx 3.0 \, \mu {\rm eV}$. 
 Note that this energy difference has to be 
 calculated quite carefully within the Gutzwiller 
 approach. In particular, one has to keep in mind that any approximation
 on the parameters  $\lambda_{\Gamma,\Gamma'}$ that reduces the variational 
flexibility may lead to a grossly 
overestimated anisotropy energy. This is a serious problem, in particular,
 in the case of Iron. For Nickel, however, the 
active multiplet states belong mostly to $d^8$ and $d^9$. Here, 
a mixing of states $|\Gamma\rangle,|\Gamma'\rangle$ has little effect 
on the variational energy, and even a diagonal variational parameter
 matrix $\lambda_{\Gamma,\Gamma'}\sim \delta_{\Gamma,\Gamma'}$ would lead to 
reasonable results.

\begin{figure*}[htb]
\begin{center}
\includegraphics[scale=0.9]{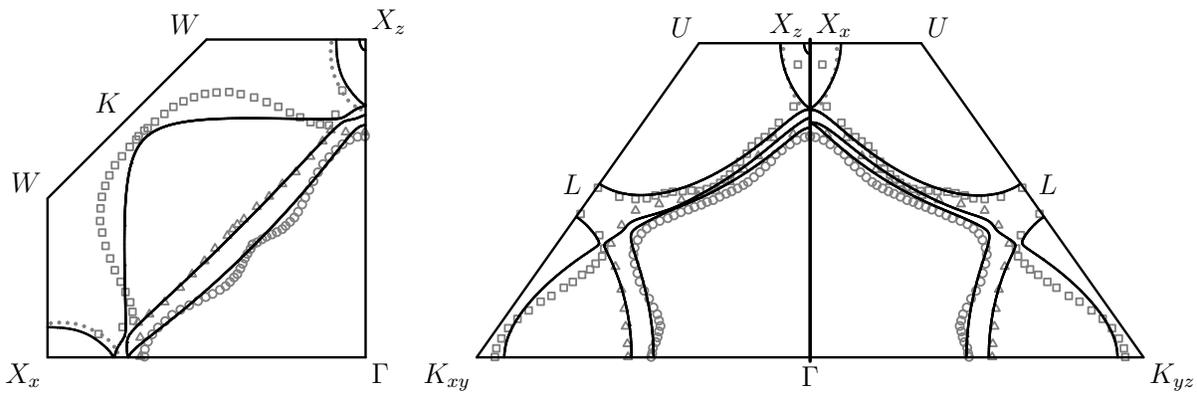}
\caption{Fermi-surface cuts with various planes in the Brillouin zone.
Lines: Gutzwiller theory including spin-orbit coupling;
Squares and triangles: experimental data reported 
in~\protect\onlinecite{stark};
Dots: experimental data of Tsui~\protect\cite{tsui}.
\label{fermi-surface-cut}}
\end{center}
\end{figure*}

In Figures \ref{fig1} and \ref{fig2} we show the quasi-particle 
band-structure that 
 arises from our calculation around the 
 $X$-points $X_{z}\equiv(001)$ and $X_{x}\equiv(100)$. 
When the magnetic moment 
is along  the easy axis, the band-structure around both 
$X$-points  coincides and the minority state $X_{2\downarrow}$ is below the 
Fermi-energy.\cite{x2down} For a magnetic moment along the $[001]$-direction, 
however, 
 the two states $X_{2\downarrow}$ have different energies. The 
$X_{2\downarrow}$ state at $X_{x}$ remains below the Fermi level, 
 whereas the corresponding state at $X_{z}$ creates a new hole pocket 
around this $X$-point. 
This is the scenario proposed by Gersdorf.\cite{gersdorf2}

 In Figure~\ref{fermi-surface-cut} 
we show Fermi-surface 
cuts that we find within our Gutzwiller theory. The experimental values are 
taken from dHvA experiments by Tsui~\cite{tsui} and by
Stark as reported in Ref.~\onlinecite{stark}. 
The agreement is quite satisfactory along
 high-symmetry lines, whereas there are significant discrepancies away 
 from them. We do believe that the wiggles that appear in the 
 experimental data are, in fact, spurious. The derivation of a Fermi surface 
from the raw dHvA data requires an expansion in Fermi surface harmonics, with 
the coefficients of the harmonics to be determined form least squares' fits 
to the data. Possibly, an over-determination occurred 
which led to unphysically 
large higher harmonics coefficients and resulted in the wiggles. 
We propose to redo these measurements. 

  In summary, we have resolved the long-standing problem to 
 explain theoretically the electronic and magnetic properties of 
elementary fcc-Nickel. Our calculations are based on the 
 Gutzwiller variational theory which is a powerful tool 
 for the investigation of Fermi-liquid systems with medium to strong 
Coulomb interaction. For such systems, state of the art band-structure
 theories  usually fail. 
Our results for the quasi-particle bands are in very good 
agreement with ARPES experiments and we find the experimental Fermi-surface 
topology. Furthermore, we are able to explain the subtle effects that 
 the spin-orbit coupling has in Nickel. Our theory yields the correct 
 anisotropy energy and we confirm the Gersdorf scenario:
 The Fermi-surface topology changes around the $X$-point $(001)$ when the 
 magnetic moment direction is rotated from $\vec{\mu}||[111]$ to 
$\vec{\mu}||[001]$ by an external magnetic field.


\begin{thebibliography}{22}
\bibitem{comment1}In the presence of spin-orbit coupling there may also 
be a  finite parameter $\eta_{\sigma,\sigma'}$ for a coupling between
 the $s$-orbital and a certain $d$-state. However, we found that this 
 term does not affect the results for Nickel significantly.
\bibitem{x2down}Note that in the presence of spin-orbit  coupling 
the spin label in $X_{2\downarrow}$ is not well defined but merely 
 refers to the main spin-component of this state.
\end{thebibliography}
\end{document}